\documentclass{article}
\usepackage{sao2,psfig}
\usepackage{graphicx}
\usepackage{amssymb}
\usepackage{url}
\issue{2003, 56, 5-14}
\begin{document}
\title{Earliest spectroscopy of the GRB~030329 optical transient}
\author{V.V. Sokolov\inst{a}\thanks{Email: sokolov@sao.ru (VVS),
          timur@sao.ru (TAF)} 
          \and T.A. Fatkhullin\inst{a}
	  \and V.N. Komarova\inst{a} 
          \and V.G. Kurt\inst{c}
	  \and V.S. Lebedev\inst{a}
	  \and A.J. Castro-Tirado\inst{b}
	  \and S. Guziy\inst{d,b} 
          \and J. Gorosabel\inst{b,f}
          \and A. de Ugarte Postigo\inst{b}
	  \and A.M. Cherepaschuk\inst{e}
	  \and K.A. Postnov\inst{e}
}
\institute{
\saoname
\and
Instituto de Astrof\'{\i}sica de Andaluc\'{\i}a
(IAA-CSIC), P.O., Box 03004, E-18080 Granada, Spain
\and
Astro Space Center of the Russian Academy of Sciences, Moscow 117810,
Russia
\and Kalinenkov Astronomical Observatory, Nikolaev State University, Nikolskaya 24, Nikolaev, 54030, Ukraine
\and Sternberg State Astronomical Institute, Universitetskii pr. 13, 119992 Moscow, Russia
\and Space Telescope Science Institute, 3700 San Martin Drive,
Baltimore, MD 21218 USA
}
\date{June 26, 2003}{November 28, 2003}
\maketitle
\begin{abstract}
We present the earliest BTA (SAO RAS 6-m telescope) spectroscopic observations
of the GRB~030329 optical transient (OT), which almost coincide
in time with the first ``break'' ($t\sim 0.5$ day after the GRB) of the OT
light curve. The BTA/MPFS (Multi-Pupil Fiber Spectrograph) spectra
are clearly not smooth: the OT spectra
showed a continuum with several broad spectral features
at about $4000,\ 4450,\ 5900$\AA.
The OT spectrum showed some evolution
starting from the first night after the burst,
and the beginning of spectral changes are seen as early as
$\sim 10-12$ hours after the GRB.
The onset of the spectral changes for $t < 1$ day indicates
that the contribution from Type Ic supernova (SN) into
the OT optical flux can be detected earlier.
So, summarizing our and other spectroscopic data on the OT
we confirm the evolution of the OT spectrum and color
pointed also by other authors (Hjorth et al. 2003; Matheson et al. 2003b).
The properties of early spectra of GRB~030329/SN~2003dh
 can be consistent with
a shock moving into a stellar wind formed
from the pre-SN.
Such a behavior (similar to that near the UV shock breakout in SNe)
can be explained by the existence
of a dense matter in the immediate surroundings of GRB/SN progenitor.

\end{abstract}

\section{Introduction}
Observations of host galaxies and
transient sources related to gamma-ray bursts (GRBs)
in wide spectral range from X-ray to radio
lend growing credence to the
association of GRB with certain types of supernova (SN) explosions.
It seems the most significant and important result
obtained during the last six years.
So-called red bumps (there are already more than 10 events)
detected in light curves of GRB optical transient (OT)
are the first evidence of GRB and SN connection
(see e.g. Galama et al. 2000; Bloom et al. 1999;
Castro-Tirado and Gorosabel 1999; Castro-Tirado et al. 2001 ;
Sokolov 2001b; Lazzati et al. 2001; Bjornsson et al. 2001;
Bloom et al. 2002).
Time variation in the light curves of GRB OTs
as well as emission and absoption lines in X-ray afterglows
(Amati et al. 2000; Piro et al. 2000)
is an indication to a dense matter in the vicinity of massive progenitor stars.
Such circumburst medium (or density inhomogeneities)
arising naturally due to stellar winds emanating from the massive stars
is a result of the evolution of massive stars
which are progenitors of core-collapse SNe (Chevalier 2003a).

Finally, the intensive massive star formation in GRB host galaxies
gave the very first evidence of the link between GRBs and massive stars
(Djorgovski et al. 2001; Sokolov et al. 2001c; Chary et al. 2002; Le Floc'h et al. 2003).
But, it is clear that a crucial experiment confirming the relation between
GRB and core-collapse SNe is the search for and further accumulation
of ever increasing information on SN-like features in spectra of
GRB OTs, starting from the earliest
changes in the objests' luminosities.

A long-duration and extremely bright gamma-ray burst (GRB~030329) triggered by
the FREGATE instrument on board the
HETE-II satellite  on 29 March 2003 at 11:37:14.67 UTC ($t_0=$March 29.4841 UT).
Ground-based analysis of the HETE-II SXC data localized
this event within an error box of
2 arcmin radius centered at $\alpha_{J2000} = 10^h44^m50^s$ and
$\delta_{J2000} = +21\degr30\arcmin54\arcsec$ (Vanderspek et al. 2003).
The burst duration in the $30-400$ keV band was > 25 s.
The fluence  of the burst was $\sim 1 \times 10^{-4} ergs\,cm^{-2}$
and the peak flux over 1.2 s was
$ > 7\times10^{-6} ergs\,cm^{-2}\,s^{-1}$ (i.e., $> 100\times$ Crab flux) in the same energy
band (Vanderspek et al. 2003).
The OT was detected as a bright new object of $R\approx12\fm4$
(Peterson \& Price 2003; Torii 2003; Price \& Peterson 2003).
The OT position was measured as follows:
$\alpha_{J2000} = 10^h44^m49\fs957$ and
$\delta_{J2000} = +21\degr31\arcmin17\farcs46$ (Henden 2003).
High resolution spectroscopy with the UVES instrument at the VLT
unit Kueyen revealed emission lines which were identified as $H_\beta$, [OIII] doublet and $H_\alpha$ at
redshift $z=0.1685$ (Greiner et al. 2003). It should be noted that this event is the nearest GRB
with measured redshift to date
(may be with the exception of the peculiar GRB 980425).

In this paper we present the earliest optical spectra of GRB 030329
OT obtained with the BTA (the 6 meter telescope of SAO RAS)
within 10.8 to 12.4 hours after the burst.
Sections 2 and 3 report about observations, data processing
and comparison with the data by other authors.
Section 4 shows  the onset of the spectral changes
and early spectra of the OT and SNe.
Possible consequences of
the complicated structure of GRB 030329 OT light curves
and the earliest spectroscopy of the OT
are discussed later in Sections 4 and 5.

\section{Observations and data reduction}

Spectroscopic observations of the GRB~030329 OT were performed on 29/30 March
2003  with the Multi-Pupil Fiber Spectrograph (MPFS)
(see WWW-page at \url{http://www.sao.ru/~gafan/devices/mpfs/mpfs_main.htm})
 at the 6 meter telescope (BTA) of SAO RAS starting 10.8 hours after the burst.
The observations were carried out by A. Moiseev, researcher of the
Laboratory of Spectroscopy and Photometry of Extragalactic Objects of SAO RAS.
Photometric conditions on the night were good with slightly variable seeing.
The log of the observations is presented in Table~\ref{spec_log}.
The spectrograph was equipped with a $2048\times2048$
  EEV CCD42-40 chip and pixel size of $13.5\,\mu m$.
In the observations a 300 lines/mm grating
  blazed at 6000\AA\ was used giving a reciprocal dispersion of about 2.8\AA\
  per pixel and an effective wavelength coverage
  of 3800--9600\AA.
Spectral resolution measured by sky emission lines is about 12\AA\
  (full width at half maximum, FWHM).
Eight 600-second spectrum images were obtained in all.

The data were processed using an IDL reduction
  package written by V.L. Afanasiev.
Reduction included bias subtraction, flat-fielding, removing of cosmics,
  measuring positions of the individual spectra,
  extraction of object spectrum images, subtraction of sky spectrum,
  linearization using lines of a neon calibration lamp
  and correction for differential spectral sensitivity
  using the spectrophotometric standard Feige 56.
The linearization was inspected by measuring wavelengths of night sky lines,
  and we found that the mean error of dispersion curves was $0.2$ pixel.
Unfortunately, in the standard star spectrum the second
  spectral order appeared and real correction for differential
  spectral sensitivity was possible only to 7200\AA.
As a result, sums of two consecutive spectra were used,
  so, finally, we obtained four individual 1200 second spectra,
at the mean epochs:
  0.45 ($10\fh8$), 0.47 ($11\fh3$), 0.48 ($11\fh5$) and 0.52 ($12\fh4$)
  days after the burst, respectively.
\begin{table}
\caption{The log of spectroscopic observations of GRB 030329 OT with the BTA ($t_0 = $March 29.4841 UT)}
\label{spec_log}
\begin{center}
\resizebox{\hsize}{!}{
\begin{tabular}{lccc}
\hline\hline
Date, UT            & Exposure      & Zenithal  & Seeing \\
(start of exposure) & time (sec)    & distance  &  (FWHM) \\
\hline
March 29.9213, 2003 & 600           & $38\degr$         & $1\farcs5$\\
March 29.9300, 2003 & 600           & $40\degr$         & $1\farcs5$\\
March 29.9423, 2003 & 600           & $43\degr$         & $1\farcs5$\\
March 29.9507, 2003 & 600           & $46\degr$         & $1\farcs5$\\
March 29.9589, 2003 & 600           & $48\degr$         & $1\farcs5$\\
March 29.9674, 2003 & 600           & $50\degr$         & $1\farcs5$\\
March 29.9955, 2003 & 600           & $57\degr$         & $1\farcs8$\\
March 30.0040, 2003 & 600           & $59\degr$         & $1\farcs8$\\
\cline{2-3}
                    &  Feige 56     &                   &           \\
March 29.9826, 2003 & 20           & $47\degr$         & $1\farcs8$\\
March 29.9844, 2003 & 20           & $47\degr$         & $1\farcs8$\\
\hline\hline
\end{tabular}
}
\end{center}
\end{table}

\section{Spectroscopy}

As MPFS is a two-dimensional spectrograph,
   some circular aperture should be defined during the extraction of spectrum.
In the spectroscopic obrervations of GRB~030329 OT
   the aperture diameter of 7 lenses was used.
This corresponds to about 7 arcsec in spatial dimension,
and the same aperture size was used for all OT spectra.

In order to control the absolute flux calibration,
the photometry and spectroscopy were compared.
We used $UVR_cI_c$ photometric observations of the OT carried out
   with the Zeiss-1000 telescope
   at SAO RAS on the same epoch as the BTA/MPFS spectroscopy
   was carried out
   (the magnitudes and full $UBVR_cI_c$ light curves are reported by
   Gorosabel et al. 2003 and Guziy et al. 2003).
For the $B$ band we used the Nordic Optical Telescope observations
   (Castro-Tirado 2003).
Photometric conditions were good during the MPFS spectroscopic and
   Zeiss-1000 photometric observations.
For the spectrum obtained 12.4 hours after the GRB
   it is possible to compare directly the photometric $BVR_c$ points
   with the spectrum.
Using the OT magnitudes from Castro-Tirado et al. (2003)
   and zero points from Fukugita et al. (1995), we performed this comparison.
The spectral range of the obtained spectra allows us to compare only
   $B$ and $V$ broad-band fluxes.
Thus, the spectrum was
   integrated with $B$ and $V$ standard filter transmission curves.
We found that the spectrum
   should be corrected by a factor of $0.89$ in fluxes for coincidence
   with the photometry that is fully within
the uncertainty of spectrum image extraction.
As can be seen in Figure~\ref{spec-phot},
   the photometry and spectroscopy
   are in good agreement after the correction.
This comparison allows us to be sure that the reduction for differential spectral
   sensitivity was correct.
As we used the same sensitivity curve for all spectra,
   we can scale them by the same factor.

\begin{figure*}[th]
\vbox{
\centerline{\includegraphics[width=0.8\textwidth,clip]{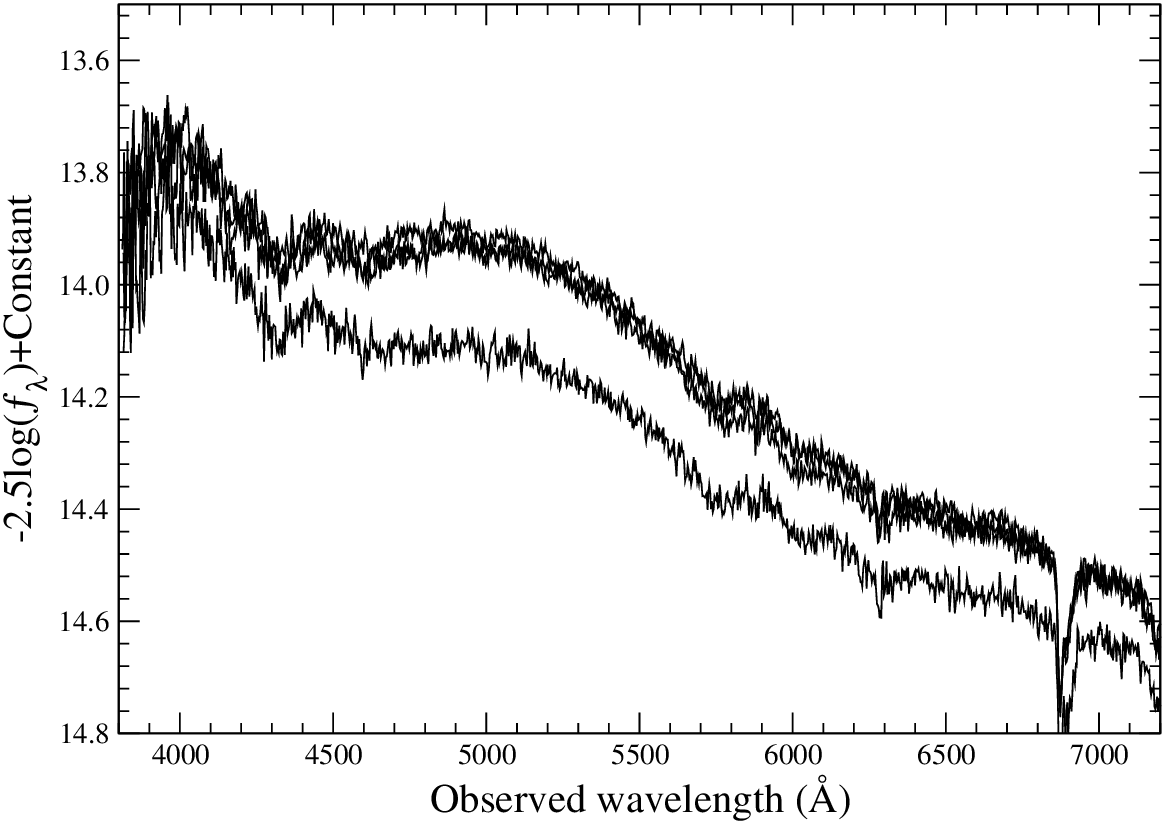}}
\par
\caption{The four BTA/MPFS spectra of GRB~030329 OT
are presented in the order of flux decrease
for 0.45 ($10\fh8$), 0.47 ($11\fh3$), 0.48 ($11\fh5$) and 0.52 ($12\fh4$)
days after the burst, respectively.
The $10\fh8$, $11\fh3$, and $11\fh5$ spectra
correspond to almost equal fluxes (see Figure~\ref{filt_spec}).
}\label{spectra}

\vspace{5mm}

\centerline{\includegraphics[width=0.8\textwidth,clip]{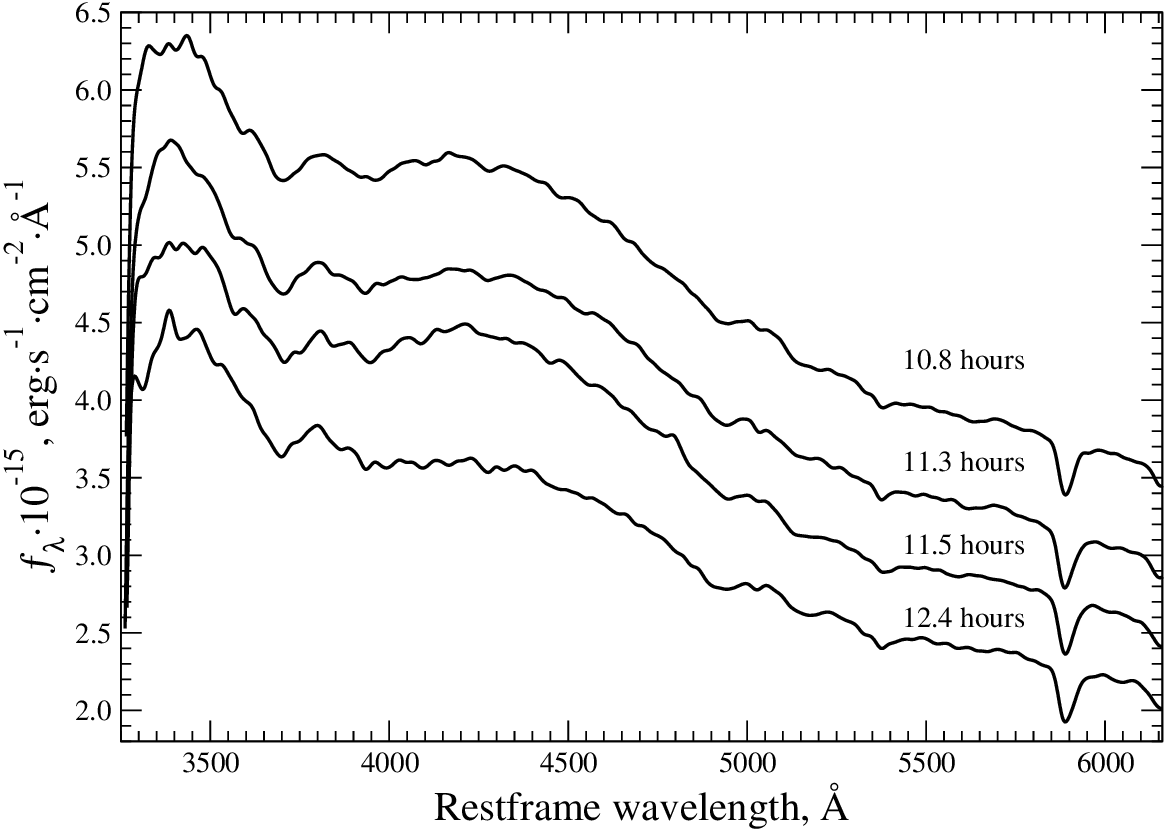}}
\caption{The MPFS spectra (in restframe wavelengths) smoothed by a gaussian
with FWHM equal to MPFS spectral resolution (12\AA).
The smoothed spectra of GRB~030329 OT
were shifted up the scale of $f_\lambda$
relative to the last (12.4 h) spectrum by
+0.2E-15 for 11.5 h, +0.6E-15 for 11.3 h, and +1.2E-15 for 10.8 h spectra,
respectively.}\label{filt_spec}
}
\end{figure*}

\begin{figure*}
\centerline{\includegraphics[width=0.8\textwidth,clip]{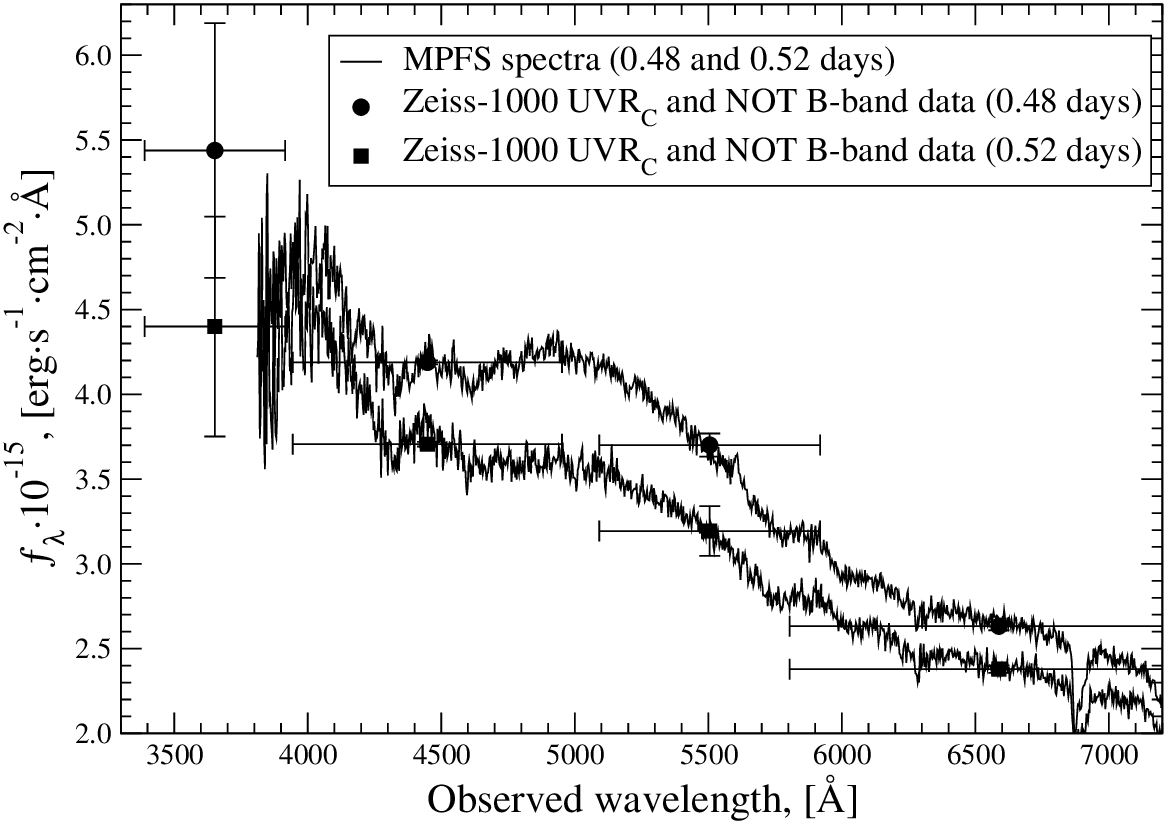}}
\par
\caption{Comparison of spectroscopy and photometry.
By horizontal bars the FWHM (full width at half maximum)
of corresponding filters are shown.}\label{spec-phot}

\vspace{5mm}

\centerline{\includegraphics[width=0.82\textwidth,clip]{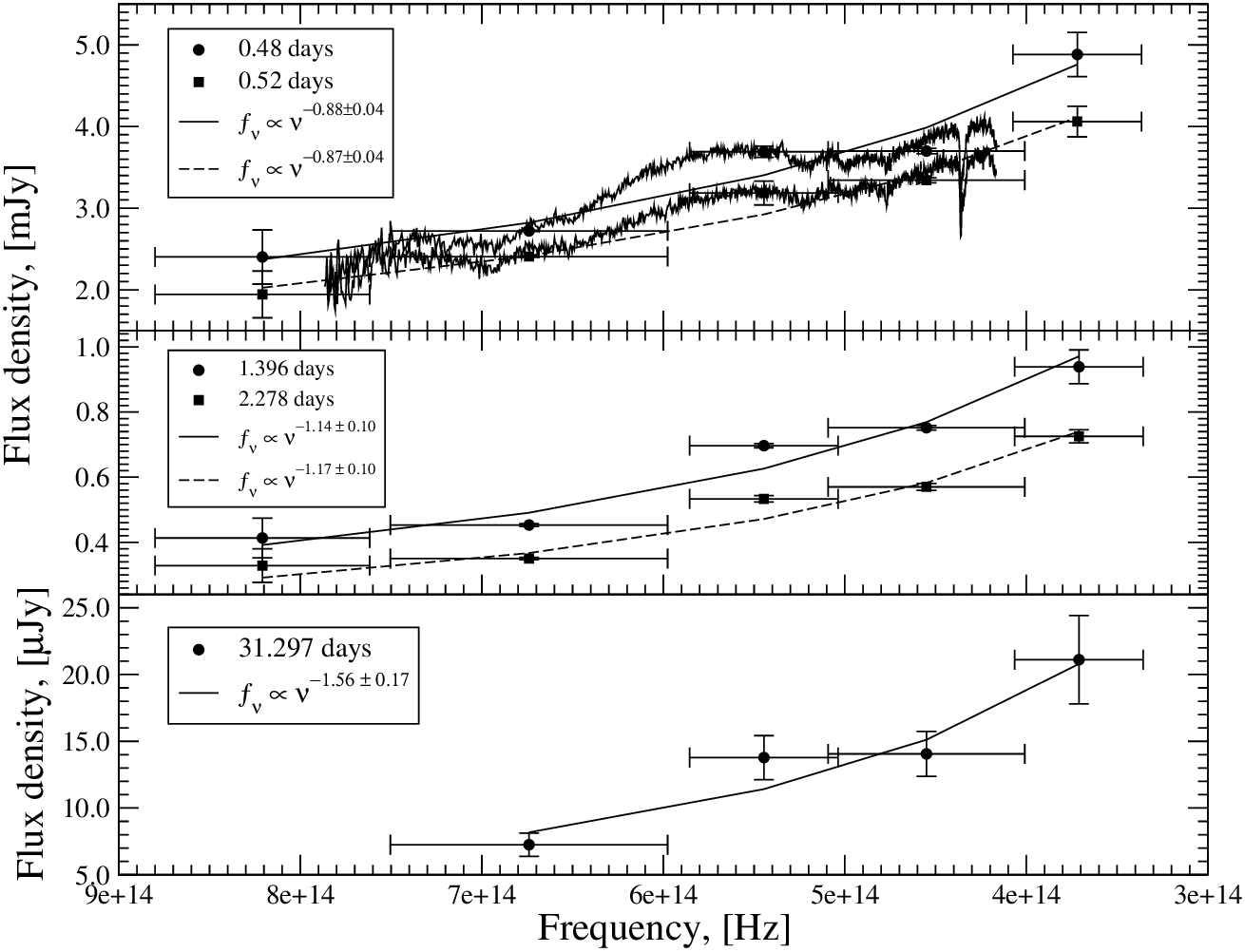}}
\caption{Evolution of the $UBVR_cI_c$ broad-band spectra
during the first three nights. The MPFS spectra are also shown.
It is clearly seen from the figures that systematic deviation of V-band flux
from formal smooth power-law is due to real unsmooth OT spectra.
Also from these figures one can conclude that broad spectral features
remained significant during the first three nights.}\label{spec_evol}
\end{figure*}

Figure~\ref{spectra} presents the resulting spectra of the GRB~030329 OT.
The four spectra are presented in the order of flux decrease
for 0.45 ($10\fh8$), 0.47 ($11\fh3$), 0.48 ($11\fh5$) and 0.52 ($12\fh4$)
days after the burst, respectively.
The $10\fh8$, $11\fh3$, and $11\fh5$ spectra
correspond to almost equal fluxes,
that is why in Figure~\ref{filt_spec} they are shifted with respect to the last
spectrum (for $12\fh4$)
with arbitrary constants.
In Table~\ref{sn_est} the signal-to-noise ratios for the resulting
   spectra are presented.
Finally, it should be noted that there are
   no strong sky emission bands within the range of the spectra.
Thus, broad spectral features (Figures~\ref{spectra} and \ref{filt_spec}) 
detected in the spectra are confidently real.

\begin{table}
\caption{The signal-to-noise ratios for MPFS spectra of GRB 030329 OT}
\label{sn_est}
\begin{center}
\begin{tabular}{cccc}
\hline\hline
Epoch, & S/N      & S/N      & S/N      \\
days   & 4000\AA\ & 5000\AA\ & 7200\AA\ \\
\hline
0.45   &   31     &  100     &  48 \\
0.47   &   25     &  99      &  52 \\
0.48   &   23     &  93      &  50 \\
0.52   &   17     &  60      &  44 \\
\hline\hline
\end{tabular}
\end{center}
\end{table}

\section{Disscusion}

First of all it should be noted that our spectra
are the earliest OT spectra obtained for the GRB~030329 event.
As can be seen from Figure~\ref{spectra} (and from Figure~\ref{filt_spec})
the spectra showed
an unsmooth continuum with several broad spectral features
at about $4000,\ 4450,\ 5900$\AA.
A comprehensive interpretation of these details is a subject
for a future study, therefore, for simplicity we assumed here
that central wavelengths correspond to local maxima of the details.
Broad spectral features similar to these ones
are characteristic of GRB 030329 OT spectra:
compare, for instance, our spectra in Fig.\,1 or 3 with the observed spectrum of
GRB 030329/SN 2003dh in Fig.\,10 from the paper by Matheson et al. (2003b).
It is interesting that the following spectral observations by other groups
showed an evolution of these features.
No spectral features were detected on UT 1--3 April spectra,
but they developed again in spectra starting from UT 6 April
(Stanek et al. 2003; Hjorth et al. 2003) and continued to be detected
in spectra on UT 8 and 9 May (Kawabata et al. 2003).
Thus, ours and other data clearly indicated  the evolution of GRB~030329 OT
spectrum.

Figure~\ref{spec_evol} presents formal power-law fits of the $UBVR_cI_c$
broad-band spectra and their evolution during the first three nights.
Power-law slopes are corrected for galactic extinction $E(B-V)=0.025$
(Schlegel et al. 1998).
The MPFS spectra for the first night ($t = 0.48$ and 0.52 days after the GRB) 
are also shown.
We note, however, that as can be seen clearly in Figure~\ref{spectra},
the MPFS spectra are not smooth and there is an uncertainty
in the $UBVR_cI_c$ broad-band spectra fits.
It is clearly seen from the figures that systematic deviation of the V-band flux
from formal smooth power-law is due to real unsmoothness of the OT spectra.
From Figure~\ref{spec_evol} one can also assume that broad spectral features
remained significant during at least the first three nights
(up to $t = 2.278$ days).

The power-law slopes for our $UBVR_cI_c$ data from the Zeiss-1000 telescope
on the second and third nights
(see Fig. 4 for $t = 1.396$ and 2.278 days) are consistent
with the value $\beta \sim -1.2\pm0.1,\ f_\nu \sim \nu^\beta$
obtained directly from the OT spectrum
of UT 3.10 April ($t = 3.62$ days, see Hjorth et al. 2003),
but within only $2\sigma$ with the value
obtained from $ugriz$-band data on UT 31 March
(see Lee et al. 2003).

Figure~\ref{spec_evol} shows that there is some evidence of reddening
in the OT broad-band spectrum of the first three nights.
Moreover, it should be noted that within about a month after the burst
the colors of the OT
are redder than during the first three days,
which is, in turn, consistent with continuing reddening
of the broad-band spectrum. Such a behavior of
the broad-band spectrum can be explained by an increase in
SN fraction in the GRB OT light. Indeed, the {\it late} SN spectra
are considerably redder than GRB optical transient spectra
(Hjorth et al. 2003).

\subsection{On the onset of the spectral changes of GRB 030329 OT
and the earliest supernovae spectra}

All later spectra have been published
  by Stanek et al. (2003), Hjorth et al. (2003), Matheson et al. (2003b),
   Kawabata et al. (2003),
   and all necessary references are also adduced there.
In the opinion of Matheson et al. (2003b)
the optical spectroscopy (combined with the photometry)
from a variety of telescopes allows an unambiguous separation
between the GRB afterglow and SN contributions, and the transition
between the afterglow and the SN spectra is gradual.
But, as follows from our observations,
the earliest $UBVR_cI_c$ spectrum can be ``smooth'' only in appearance,
and the early evolution of the OT spectrum is possible.
After all, Matheson et al. (2003b) also directly point
to color variations during the first week.
Apart from that,
we  obtained our earlier BTA/MPFS spectra of the OT
just during the most rapid variations
in the OT luminosity and physical conditions in the source.
Below it is shown that this phase is like some SNe (1993J, 1997A)
observed during the first light curves UV peaks (or during the UV
breakout phase),
specially by their similar fast luminosity variations,
spectra, and bolometric luminosities.
The bolometric luminosities in the first SNe UV peaks
can be also approximately of the same values as in the GRB OTs.

Concerning a possible early color/spectral variability
Matheson et al. (2003b) noted that the optical afterglow of the GRB
is initially a power-law continuum but shows significant color variations
during the first week which are unrelated to the presence of a supernova,
implying that an evolution of the ``smooth'' continuum slope is possible
according to their dataset. In particular, when analyzing the colors
(in their Fig.\,2.),
it is clearly seen that on their first night the object was indeed bluer.
It means that it is below the solid lines indicating the color
expected for an afterglow with a fixed power-law spectrum.
Matheson et al. (2003b) themselves note
that there is a clear color change in the afterglow
over the first few days.
It is consistent with the results of our earlier BTA/MPFS spectral data.
So, judging by the spectra (and photometry by Gorosabel et al. 2003),
in the first hours the object was even bluer.
It follows also from independent $BVRI$ photometry during the first
hours (0.25 days) after the GRB: Burenin et al. (2003) observed
a somewhat flatter ($\beta = -0.66$) or bluer spectrum than
Matheson et al. (2003b)
($\beta = -0.97$) on the third night after the burst.

Then Matheson et al. (2003b) note
that they do not see any contribution of a supernova component
(within the uncertainties in their fits) in their early spectra.
At the same time, another observational group with VLT (Hjorth et al. 2003)
claims that the SN is apparent on April 3 UT,
i.e.  about 4 days after the GRB.
Besides, the group of Hjorth et al.  does not divide
the spectrum into two components strictly by means of a ``standard''
GRB OT spectrum, noting
that the resulting overall spectral shape of the SN contribution
does not depend on the adopted power law index or template spectrum.
This is a more cautious approach,
since such a division itself contains a lot of arbitrariness.
A question arises at once:
which spectrum of GRB afterglow can be considered the ``standard'' one?
If SNe and GRBs are  indeed produced by the same astrophysical cauldron
(Kawabata et al. 2003), then most probably
the spectra of SN and the GRB afterglow can be mixed so closely
that it would be just dangerous to divide them in the earliest stages
of the most rapid changes in the source.
It is natural to assume that at the very beginning of the GRB/SN explosion
(or in ``the rise time of SNe'', or in the SN onset time)
the contributions of {\it early} spectra of the SN
and the spectrum of the GRB afterglow can change quickly
into the common (observable) spectrum of the GRB OT.
Thus, the relative SN/OT contribution to the earliest integrated spectra might
be rapidly variable.

At some moment these contributions
can become even comparable in bolometric luminosities
(as an example, see estimations of bolometric luminosity in the first maximum
of SN 1993J from Shigeyama et al. 1994).
This is reinforced by the fact that the earliest spectra of SNe
are very similar to the GRB afterglow spectra in their powerful UV continuum.
And especially since (as Matheson et al. note for SNe Ic)
the rise time (more exactly, the beginning/onset time of the explosion)
of the majority of known SNe are not well defined
(see Norris \& Bonnell (2003) for more detailed discussion of this problem.)
As an illustration, we show in Fig. 5
the examples of such early spectra corresponding to the UV shock breakout in
two core collapse SNe (SN 1993J, SN 1987A),
which have the most exactly defined times of the explosion onset.

SN 1993J may be the only event for which good data are available
(except for SN 1987A), and there were also spectroscopic indications
that the SN 1993J was undergoing the transition from Type II to Type Ib
(Filippenko \& Matheson 1993). The early spectra of SN 1993J showed
(Filippenko \& Matheson 2003) an almost featureless blue continuum up to 
$5000$\AA, possibly, with broad ($>5000$\AA), but weak
hydrogen and helium lines ($H_\alpha$, HeI $5876$\AA\ ...).
Both the spectra and the light curves of SN 1993J indicate
that it was not a typical Type II SN.
SN 1993J rose quickly to the first luminosity maximum and
then rapidly declined during $\sim 1$ week,
only to brighten a second time over the following two weeks.
At the very beginning
SN 1993J had  strong UV excess with a relatively smooth
spectrum up to $5000$\AA\ (see Fig. 5).
Initially unusual light curves and
the appearance of the He I lines in
the spectra were interpreted as
evidence that SN 1993J was similar
to a SN Ib,
with a low-mass outer layer of
hydrogen (that gave the early
impression of a SN II).
It can be said that its emission comes from
the collision of supernova ejecta
with circumstellar gas
that was released by the progenitor star
prior to the explosion (Chevalier 2003b).
Such a  mass loss is consistent with
the fact that the SN properties
indicate that most of the stellar H
envelope is present at the time of the
SN explosion.
  SN 1987A is another core collapse supernova that exploded
with a (massive) H envelope.
But the immediate surrounding of the progenitor star
(the pre-supernova is a blue supergiant)
was determined by the fast wind from that star
      (Chevalier 2003b).

In a sense, the famous SN 1993J IIb is indeed a SN of a transition class
to the Type Ib/c SNe with a ``bare'' Wolf-Rayet (WR) pre-SNe,
which can be GRB progenitors (more exactly pre-GRBs).
In SNe IIb (evolve to Ib/c) most of the H envelope
is lost before the SN explosion.
The Types Ib/Ic are core collapse SNe, which are thought to have massive
star progenitors that lost their H envelopes.
In the latter case, no slow H envelope is expected in the SN,
and the {\it immediate} surroundings of the SN
are determined by the low density fast wind typical of a WR star.
But the main thing that should be emphasized here
is that the variety of envelopes surrounding pre-SNe is quite natural
in the evolution of a massive star (see  Chevalier 2003a for details).
  For the interpretation of the OT light curves and spectra it should be
  emphasized first of all,
  that SNe are not constant objects with fixed standard spectra.
  Both their brightnesses and spectra change.
So-called ``typical'' SN spectra of (usually)
long-lived luminosity peaks are known.
But which is the very first SN spectrum
in the phase of its most rapid variability?
(Most probably it is this phase that is related to the GRB/SN phenomenon.)
When can the GRB afterglow spectrum  be compared
to the SN one and subtracted from the integrated GRB/SN spectrum?

It is important for us here to emphasize also
that the bolometric luminosity of SN 1993J reaches the first maximum
(according to different model estimates) of order of
    $\sim10^{45}$\,ergs\,$s^{-1}$
     $4 - 5$ hours after the core collapse
    or $\approx$ {\it onset time} of the SN (Shigeyama et al. 1994).
This luminosity is approximately equal to that of GRB~030329 OT
at the moment when (at $\sim 11$ h)
we obtained spectra with the BTA.
Also in the models of SN 1993J by Shigeyama et al. (1994)
the time of 4--5\,h after the core collapse is just the onset of SN 1993J.

So, in view of the assumption by Matheson et al. (2003b)
that the spectrum of the afterglow did not evolve since $\Delta$T = 5.64 days
it is natural to consider that {\it before} this date
the spectral slope (and spectrum) could change indeed.
Our earliest spectroscopy and photometry of GRB~030329 OT
together with the data by other authors also confirm it.
But the onset of the GRB/SN burst still remains unclear.
As was noticed by Matheson et al. (2003b),
color changes are apparent in the early stages of the afterglow,
even before the SN component begins to make a contribution.
These color changes have been seen
in other afterglows (Matheson et al. 2003a; Bersier et al. 2003),
but the physical mechanism that produces them is still a mystery.

\subsection{On the complicated structure of the GRB OT light curves
     and the earliest spectroscopy of GRB~030329 OT}

As noted in a report by Chevalier (2003a),
the burst GRB~030329, which was clearly
associated with Type~Ic SN, showed a break
in the light curve at $0.4-0.5$ day, which
was preliminarily interpreted as a jet break (Berger
et al. 2003; Granot \& Kumar 2003). However, it has become clear that there is
a considerable variable structure in the
light curve of the GRB OT over the first 10
days (Granot et al. 2003)
and the identification of the early jet break
cannot be made with certainty.
Since the surroundings of massive stars are
expected to be shaped by the winds emanating
from the massive progenitor stars, it could
be supposed that the physical nature of the
first/early break in the optical light curve
of the GRB~030329 OT is the same as one of all other
''bumps'' in the light curve during the first 10
days. In principle, it can refer to all OT GRB
light curve breaks:

1) e.g. in the case of GRB~970508 OT with
its famous bump in the optical light curve at
the age of 1 day (Kopylov et al. 1997; Sokolov et al. 1998);

2) and in the case of GRB~000301c OT, when
in 3 days after the GRB some increase was also
observed in the optical afterglow lightcurve
(Masetti et al. 2000; Panaitescu 2001, and references therein);

3) and in the case of GRB~021004 OT whose
light curves also showed a variability
superposed on the overall trend (Holland
et al. 2003; Heyl \& Perna 2003);

4) and, probably, in the case of
GRB~030226 OT with a possible bump in the R
light curve on the first day (Dai \& Wu, 2003).

This increasing list of GRB OTs with
the enigmatic/mysterious/strange
early variability in the light curves
allows us to suppose in general
that all observable early ``jet'' breaks for
other OT GRBs (with smooth light curves)
may be just due to density inhomogeneities
in the circumburst medium arising naturally
as a result of a massive star (or the GRB/SN
progenitor) evolution.
The winds of WR stars must be inhomogeneous
(Ramirez-Ruiz et al. 2001).
But the observed degree of inhomogeneity
refers to a region close to the stellar surface
(Chevalier 2003a) or to the immediate
surroundings of GRB/SN. The inhomogeneity may
decrease in the outer parts of the WR wind
where the GRB afterglow just occurs under
the ``standard'' theory (Kumar \& Panaitescu 2000;
Panaitescu \& Kumar 2002; Granot \& Kumar 2003).

Yet, let us assume that
the occurrence of a GRB should not be
associated with an internal (or external) shock
wave, but directly with the core collapse of a massive
star progenitor (or the pre-GRB/SN). It might indicate that a
long-duration GRB itself must arise in a very small
volume of a typical size of $\sim10^9$\,cm (and less)
and the associated GRB afterglow occurs at $r \lesssim 10^{15}$\,cm
(Sokolov 2001a).
At least
such an idea of the GRB/SN origin has no problems
with the explanation of unnaturally small medium
densities, which are to be ascribed
to typical starburst regions
(Chevalier 2003a) according to the
``standard/mainstream'' models of GRB origin.

But if, in fact, GRBs are always and immediately/directly related to
 core collapse SNe, then this assumption
corresponds to a considerably lower total GRB
energy ($\sim 10^{49}$\,ergs)
than has been accepted to date. And only a {\it small part}
(or a narrow collimated part) of the total GRB radiation
(probably, a few percent) reaches
the observer (Sokolov 2001b; Lamb et al. 2003).
Such (nearly spherically symmetric)
``GRB flashes'' emerge somewhere inside a compact region
(of dimension of  $\sim10^8 - 10^9$\,cm) near the collapsing core.
So, GRBs could be produced by a relativistic collapse (or an explosion)
of the massive and compact core of the progenitor star.
As this takes place relativistic jets arise concurrently
with the GRB outburst as a result
of the asymmetric explosion and the {\it jets} collimated
closely along the GRB {\it beams}.

 Then asymmetric GRB ejecta (or SN ejecta),
traveling at a nearly the speed of light,
impacts the pre-SN envelope/shell with $R \sim 10^{13} - 10^{15}$\,cm.
It is at this time or concurrently with the GRB afterglow
(if the relativistic jet is directed towards us!) when the SNe spectra
peaking at UV (like ones shown in Fig. 5) are also observed.
The shock wave related to the jets propagates
rather fast through the pre-SN envelope/shell and
arrives at the {\it stellar surface}
after the core collapse (or GRB).
However, we do not know yet detailed calculations
for such a scenario of the GRB/SN origin.

Our observations of GRB~030329 OT spectra
almost coincide in time with the first ``break''
($t \sim 0.5$ day) of the light curve
(see Fig.1 from Granot et al. 2003).
Thus, this first ``break'' of the light curve
could be interpreted as an emergence of the shock wave onto
the surface of the multilayer pre-SN star.
This is the moment when
the very first signs of the {\it earliest} SN spectrum
or deviations from the smooth GRB afterglow
power-law spectrum can be seen.
In this case the term ``stellar surface''
(Fatkhullin et al. 2003)
means the layered/complicated structure of the SN
progenitor or the structure of the massive pre-GRB/SN star.

\begin{figure*}
\begin{center}
\includegraphics[width=0.85\textwidth,bb=15 15 275 208,clip]{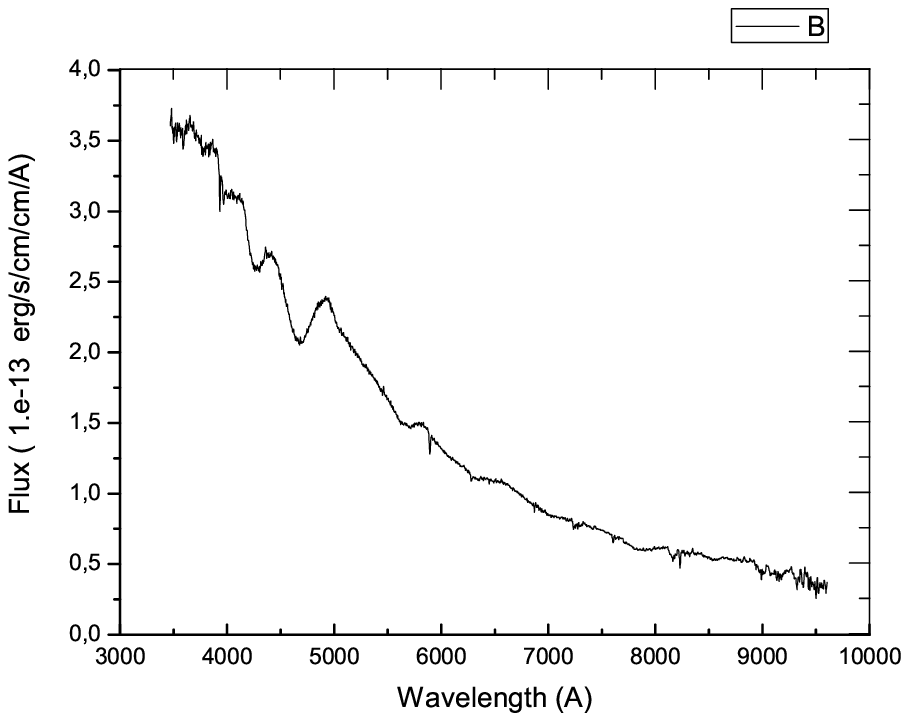}
\vspace{0.5cm}
\includegraphics[width=0.85\textwidth,bb=15 15 275 208,clip]{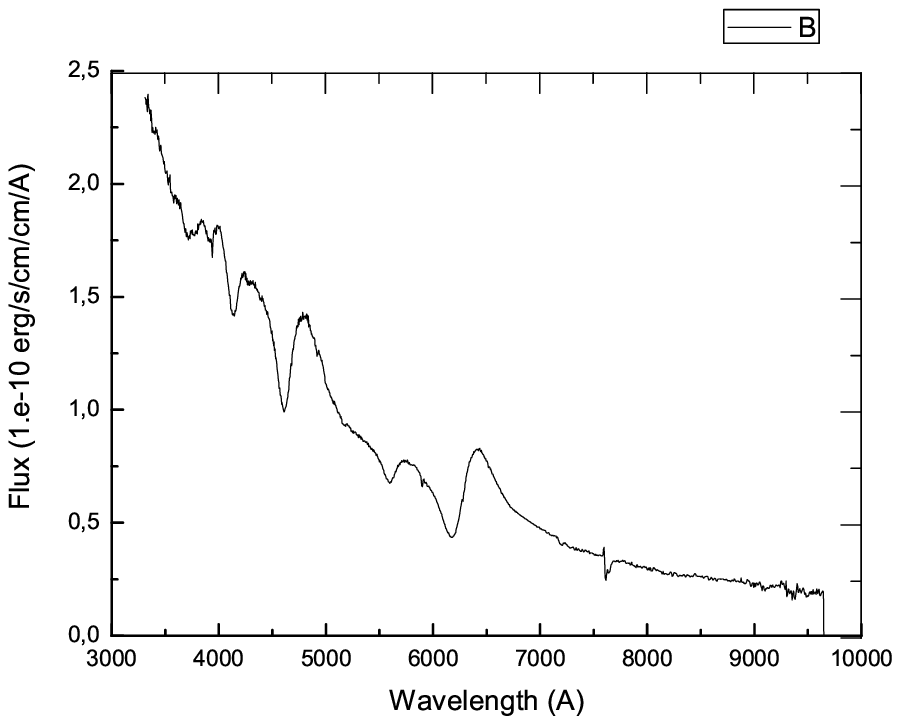}
\end{center}
\caption{
The examples of earliest optical spectra
SN 1993J and SN 1987A are given. Spectra are taken from the database
SUSPECT --- The Online Supernova Spectrum Database
http://bruford.nhn.ou.edu /~suspect/ (Richardson at al. 2002).}
\label{SN_spectra}
\end{figure*}

\section{Conclusions}
\begin{enumerate}
\item We obtained the earliest spectra of GRB~030329 OT, which
     almost coincide in time with the first ``break''
      ($t \sim 0.5$ day) of the OT GRB light curve.
\item
The BTA/MPFS spectra are clearly not smooth:
    as can be seen from Figure~\ref{spectra}
    the GRB OT spectra
    showed an unsmooth continuum
    with several broad spectral features at about $4000,\ 4450,\ 5900$\AA.
Analogous broad spectral features are characteristic of the
  GRB~030329/SN~2003dh spectra (Hjorth et al. 2003; Matheson et al. 2003b).
\item Spectra of the GRB OT showed some evolution starting from
      the first night after the burst.
      Thus, the beginning of the spectral changes
      or the systematic deviations from a smooth power-law
      is already seen
      $\sim 10-12$ hours after the GRB (Figures~\ref{filt_spec}, 3, and
      \ref{spec_evol}).
\item
  In view of the results of
   the spectral observations of GRB~030329 OT with the BTA/MPFS
     the onset of the spectral changes
       for $t < 1$ day (or on the first night)
      indicates that the contribution from Type~Ic supernova
      into the OT GRB optical flux
      can be detected earlier.
So, we confirm the evolution of the GRB OT spectrum and color,
   which are also indicated by other authors
   (Hjorth et al. 2003; Matheson et al. 2003b).
\item
The complicated structure of
GRB~030329 OT (and other GRB OTs) light curves
can be explained
as the shock propagates
through the layered structure of the GRB progenitor.
So, the structure of the massive star (pre-GRB/pre-SN) can be visible.
The early spectra of the GRB~030329 OT properties can be consistent with
a shock moving into the stellar wind formed from the pre-SN.
\end{enumerate}

\begin{acknowledgements}
Authors are grateful to A.V. Moiseev and V.L. Afanasiev for
spectroscopic observations and help in the reduction of the spectra.
This work was supported by RFBR grant No 01-02-17106.
\end{acknowledgements}

\end{document}